\documentclass[10pt,journal]{IEEEtran}
\ifCLASSINFOpdf
\else
   \usepackage[dvips]{graphicx}
\fi
\usepackage{url}
\usepackage{color}
\usepackage{booktabs}
\usepackage{multirow}
\usepackage{graphicx}
\usepackage{newtxmath}
\usepackage{amsmath}
\IEEEoverridecommandlockouts

\begin{document}
\title{Caption Feature Space Regularization  for  Audio Captioning}
\author{Yiming Zhang, Hong Yu, Ruoyi Du, Zhanyu Ma$^{\dagger}$, \IEEEmembership{Senior Member, IEEE}, Yuan Dong
\thanks{Y. Zhang, R. Du, Z. Ma, D. Yuan are with the Pattern Recognition and Intelligent System Laboratory, School of Artificial Intelligence, Beijing University of Posts and Telecommunications, Beijing 100876,
China. E-mail: \{zhangyiming, beiyoudry, mazhanyu, yuandong\}@bupt.edu.cn.}
\thanks{H.Yu is with Department of Artificial Intelligence, School of  Information and Electrical Engineering, Ludong University, Yantai, Shandong 264025, China, Email: hy@ldu.edu.cn.}
\thanks{$^{\dagger}$ The corresponding author.}}

\markboth{Journal of \LaTeX\ Class Files, Vol. XX, No. X, August }
{Shell \MakeLowercase{\textit{et al.}}: Bare Demo of IEEEtran.cls for IEEE Journals}
\maketitle

\begin{abstract}
Audio captioning aims at describing the content of audio clips with human language. Due to the ambiguity of audio, different people may perceive the same audio differently, resulting in caption disparities (\emph{i.e.}, one audio may correlate to several captions with diverse semantics). For that, general audio captioning models achieve the one-to-many training by randomly selecting a correlated caption as the ground truth for each audio. However, it leads to a significant variation in the optimization directions and weakens the model stability. To eliminate this negative effect, in this paper,  we propose a two-stage framework for audio captioning: (i) in the first stage, via the contrastive learning, we construct a proxy feature space to reduce the distances between captions correlated to the same audio, and (ii) in the second stage, the proxy feature space is utilized as additional supervision to encourage the model to be optimized in the direction that benefits all the correlated captions. We conducted extensive experiments on two datasets using four commonly used encoder and decoder architectures. Experimental results demonstrate the effectiveness of the proposed method. The code is available at \url{https://github.com/PRIS-CV/Caption-Feature-Space-Regularization}.
\end{abstract}

\begin{IEEEkeywords}
\vspace{-3pt}
Audio captioning, Contrastive learning, Cross-modal task, Caption consistency regularization
\vspace{-3pt}
\end{IEEEkeywords}
\IEEEpeerreviewmaketitle
\vspace{-7pt}
\section{Introduction}

\IEEEPARstart{a}{udio}  captioning is a cross-modal translation task that requires extracting features from audio and combining them with a language model to describe the contents of audio~\cite{drossos2017automated,drossos2020clotho,koizumi2020transformer}. However, unlike the speech recognition task that transcribes speech to text~\cite{xu2021audio}, the audio captioning task focuses on recognizing human-perceived information in general audio signals and expressing it 
with natural language. The information of generated caption includes the sound event, the acoustic scene, and some other high-level semantic information such as concepts, physical properties, and high-level knowledge~\cite{drossos2020clotho}.

\begin{figure}
\centerline{\includegraphics[width=6.5cm]{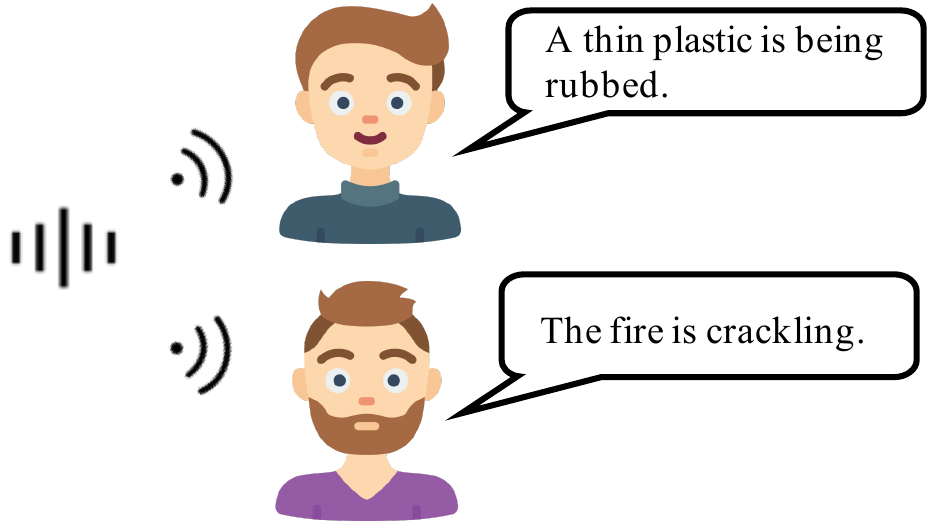}}
\vspace{-5pt}
\caption{Due to the ambiguity of audio, people may have different perceptions of the same audio.}
\label{fig:intro}
\vspace{-20pt}
\end{figure}

Different from the visual captioning tasks in which people can easily describe the visual object by its shape, color, size, and its position relative to other objects~\cite{wu2019audio}. However, for the audio clips, its information can be much more ambiguous than the information of images or videos~\cite{drossos2020clotho,wu2019audio}.
Even for people, precisely distinguishing events in audio can be difficult, let alone effectively describing the contents of given audio, because the description is often dependent on the situation or context as much as the audio itself.
Therefore, due to the ambiguity of audio, different persons may have varying perceptions of the same audio, which will result in the semantic disparity of audio captions~\cite{drossos2020clotho}, for example, a thin plastic rattling could be perceived as a fire crackling~\cite{lipping2019crowdsourcing} (as shown in Fig.~\ref{fig:intro}).

In the audio captioning task, each audio has multiple captions to describe the contents, and one caption for each training iteration is randomly selected as the training objective. While due to the disparity of captions on the same audio clip, randomly selecting one caption as the objective will lead to a large variance in the optimization direction of the model, which may harm the performance of the model and make it more difficult to converge. We believe that these captions are just semantically disparity, and the latent similarity between captions that describe the same audio clip is not aware during the current training scheme. 

Therefore, in order to solve the problem caused by the semantic disparity of captions, we propose a caption proxy space regularization method. Specifically, the proposed method includes two training stages:
(i) the first stage learns a proxy feature space of the captions by minimizing the distance between different captions belonging to the same audio and pushing away the distance between different captions of different audio, 
(ii) and the second stage trains an audio caption model with the previously built proxy space as a regularization term, i.e., utilizing the proxy feature space as an additional optimization goal to reduce optimization variance. 

\begin{figure*}
\vspace{-20pt}
\centerline{\includegraphics[width=16cm]{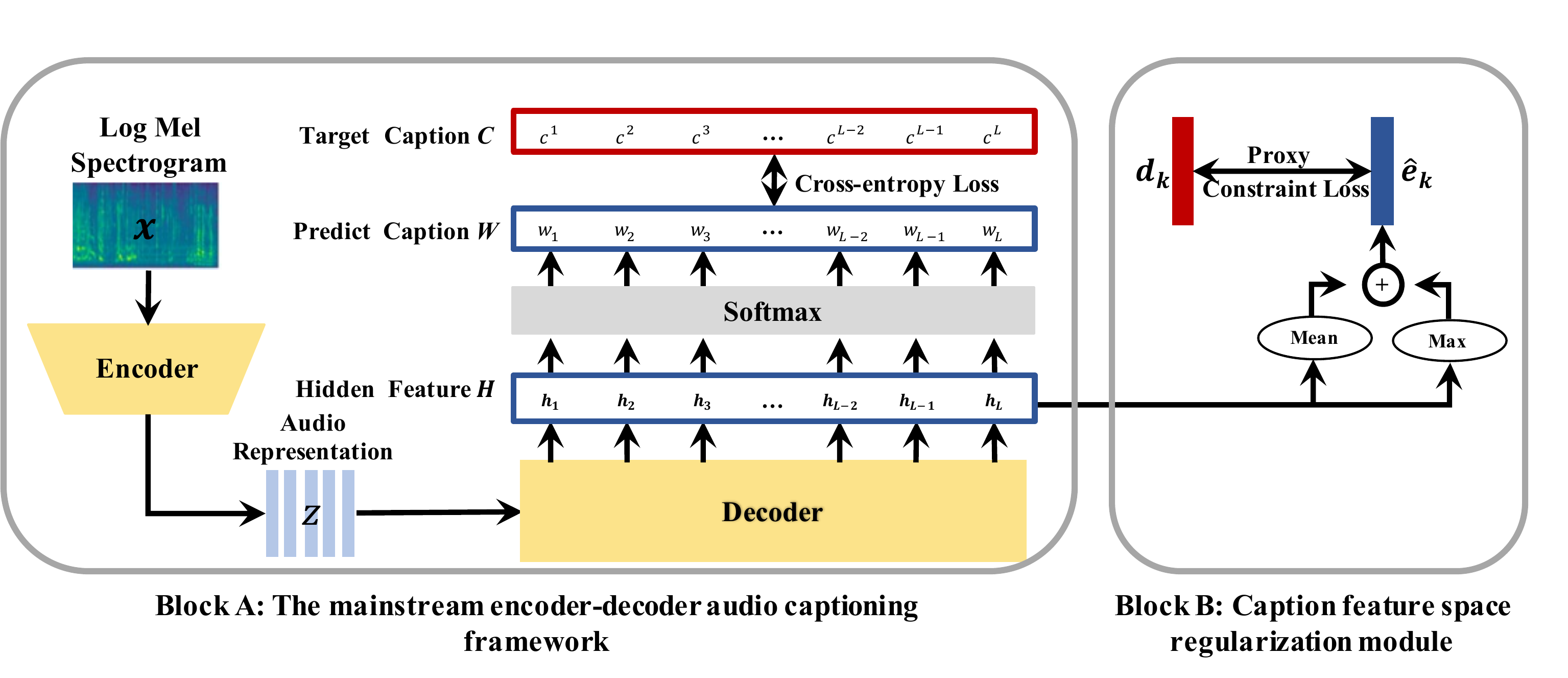}}
\vspace{-15pt}
\caption{The system overview of our proposed method. The mainstream encoder-decoder audio captioning framework is in the \textbf{Block A}, and the \textbf{Block B} is our proposed  caption feature space regularization module. $d_k$ is extracted in the first stage which is described in section~\ref{stage1} and then the $d_k$ is used to regularize the training of the caption
generation (described in section~\ref{stage2}).}
\label{fig:acmethod}
\vspace{-18pt}
\end{figure*}
Our contributions are as follows:
\begin{itemize}
  \item [1)] We propose a two-stage caption feature space regularization method to mitigate the effect of caption disparity on audio captioning task.
  \item [2)]Two different pre-trained encoders and two different decoders are used to verify the effectiveness of the method.
  \item [3)]On two prominent audio captioning datasets, Clotho-v1 and Clotho-v2~\cite{drossos2020clotho}, the results of our proposed method are compared to the results of the state-of-the-art, indicating that our proposed methods achieve competitive results.
\end{itemize}

\section{Related Work}
The audio captioning task is firstly introduced in~\cite{drossos2017automated}, which proposed the commercial ProSound Effects~\cite{lipping2019crowdsourcing} audio corpus as a proof of concept. The paper proposed a BiGRU~\cite{rana2016gated} based encoder-decoder model to generate audio captions. And due to the success of the audio captioning task in DCASE 2020 and 2021~\cite{Drossos_2019_dcase}, this task has gotten the attention of an increasing number of researchers, and several methods to address this issue have been proposed. H. Wang et al.~\cite{helinautomated} proposed a decoder with a temporal attention mechanism that uses more acoustic information for each time step. K. Chen et al.~\cite{chen2020audio} used the combination of a pre-trained encoder and a Transformer decoder which makes the latent variable result more efficient in generating captions. X. Xu et al.~\cite{xu2021investigating} investigated the effect of local and global information on the audio captioning task by comparing two pre-training tasks. The semantic information is also investigated in order to improve the audio captioning task's performance. The semantic attributes were originally used in~\cite{kim2019audiocaps}, where AudioSet labels were used as semantic attributes by using the labels of the nearest video clip. And  Eren et al.~\cite{eren2020audio} used the audio encoder to get audio embeddings and a text encoder to get subject-verb embeddings, combine these embeddings and decode them in the decoder.

\vspace{-10pt}
\section{Proposed Method}
\vspace{-1pt}\subsection{The Overview of Audio Captioning System}
\vspace{1pt}
The current mainstream training paradigm of the audio captioning task is the end-to-end encoder-decoder framework, as shown in \textbf{Block A} of Fig.~\ref{fig:acmethod}.

The training data for the task consists of paired audio and captions data. The training set of $N$ audio-captions pair $D=\left\{\left(x_{n}, \mathbb{C}_{n}\right)\right\}_{n=1}^{N}$, where $x_n\in {R}^{T \times F}$ is the log mel-spectrogram of the $n$th audio clip with $T$ frames and $F$ Mel filters, $\mathbb{C}_n=\left\{C_{mn}\right\}_{m=1}^{M}$ is all the captions of the $n$th audio clip and $C_{mn}$ is the $m$-th caption, which contains $L$ tokens $\left\{c_{mn}^l\right\}_{l=1}^{L}$.

The audio encoder takes an audio clip's log mel-spectrogram $x$ as input and extracts its latent representation $Z\in {R}^{T^{\prime} \times F^{\prime}}$. The decoder then aligns and calculates the latent representation $Z$ with the hidden states of tokens, and the caption representation can be generated by the decoder, denoted as $H \in {R}^{L \times D}$, which contains $L$ vectors $\left\{h_l\right\}_{l=1}^{L}$, where the dimension of $h_l$ is $D$ and the  number of vectors is equal to the token length of caption $C$. Hence, each vector $h_l$ corresponds to the token $c^l$ in the objective caption, the vectors are utilized to predict the probability of the words over the vocabulary after passing through the softmax layer. The predicted words are $\left\{w_l\right\}_{l=1}^{L}$.

The cross-entropy loss function is used for word-level classification of decoder, the loss function is\vspace{-5pt}
\begin{equation}
    \ell_{\mathrm{CE}}(\theta  ; C, x)=-\sum_{l=1}^{L} c^l \cdot \log p\left(w_{l} \mid \theta, x\right).
\end{equation}
\vspace{-5pt}



For the audio captioning task, an audio clip has multiple captions to describe how different annotators feel about it. Different perceptions cause the disparity of the captions in semantics, however, for each audio, only one caption will be randomly selected as the ground truth for each training to generate the description of the audio, and this training strategy can easily affect the performance of the model and make it unstable.

To solve the above problem, we propose a proxy feature optimization method to regularize the training of the caption generation, the key module is shown in \textbf{Block B} of Fig. \ref{fig:acmethod}. Our method is a two-stage audio captioning method, in which the first stage uses contrastive learning to generate the proxy space and extract the proxy caption embedding of each audio $d_k$ (described in section~\ref{stage1}) and then the proxy embedding $d_k$ is used in the module to regularize the training of the caption generation (described in section~\ref{stage2}).
\begin{figure*}
\centerline{\includegraphics[width=14cm]{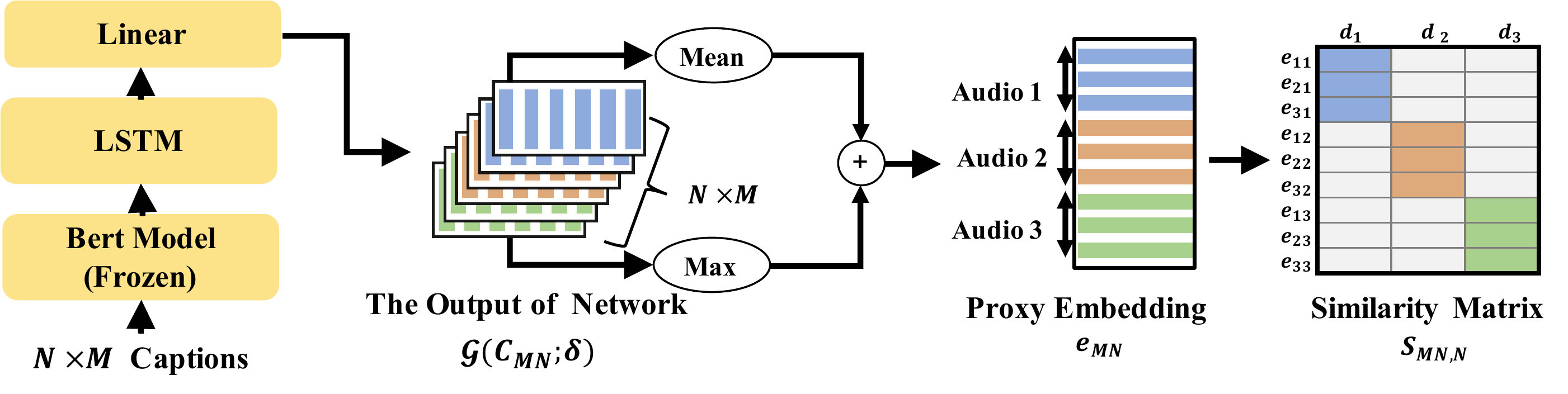}}
\vspace{-5pt}
\caption{The system overview of the first stage. Different colors indicate caption embeddings from different audio clips. The colored areas are the positive similarity and the grey areas are negative in the similarity matrix. $N$ and $M$ in this figure are 3.}
\label{fig:csembedding}
\vspace{-15pt}
\end{figure*}\vspace{-10pt}

\subsection{The First Stage : Generation of Caption Proxy Space \label{stage1}}
To reduce the effect of caption disparity on model training. We use the contrastive learning method to reduce the distance in proxy space between captions that belong to the same audio and increase the distance between captions that belong to different audio. In this subsection, we introduce the generation of the proxy space and the extraction of the caption proxy embedding.

\subsubsection{Training method}
As is depicted in Fig.~\ref{fig:csembedding}, $N \times M$ captions are fetched to build a batch. These captions are from $N$ different audio, and each audio has $M$ captions. The symbol $C_{mn}$ ($1 \le n \le N$ and $1 \le m \le M$) represents the $m$th caption from the $n$th audio clip. 

We tokenize the captions and utilize Bert~\cite{devlin2018bert} to extract its word embeddings. Then the word embeddings are fed into the LSTM network. As a network transformation, a linear layer is connected to the LSTM layer. The output of the network is $ \mathcal{G}\left ( C_{mn};\delta  \right ) $ where the parameters of the network are represented as $\delta $. The caption embedding vector is defined as
\begin{equation}
e_{mn} = \mathrm{Mean}\left ( \mathcal{G}\left ( C_{mn};\delta  \right ) \right ) + \mathrm{Max}\left ( \mathcal{G}\left ( C_{mn};\delta  \right ) \right ),\vspace{-3pt}
\end{equation}
where $e_{mn}$ represents the proxy embedding of the $n$th audio’s $m$th caption. 

The centroid of the proxy embeddings from the $k$th audio $\left [ e_{1k},\cdots ,e_{Mk} \right ] $ is denoted as $d_{k}$, and the scaled cosine similarities between each proxy embedding $e_{mn}$ to all centroids $d_{k}$ are defined by the similarity matrix $S_{MN,N}$ ($1 \le n, k \le N$ and $1 \le m \le M$):
\begin{equation}
S_{mn,k}= a \cdot \mathrm{cosine}\left ( e_{mn},d_{k} \right ) +b ,\vspace{-3pt}
\end{equation}where $a$ and $b$ are learnable parameters, the weight is limited to be positive ($a > 0$). And the centroid $d_{k}$ can be calculated by
\begin{equation}
d_{k} = \left\{\begin{array}{ll}
\frac{1}{M} \cdot \sum_{j=1}^{M}e_{j k},&if \ k \neq n  \\
\frac{1}{M-1} \cdot \sum_{j=1,j \neq m}^{M}e_{j k}, & if \ k =n
\end{array}.\right. \vspace{-3pt}
\label{eq:centroid}
\end{equation}

In Eq.~\eqref{eq:centroid}, when calculating negative similarity ($k \neq n$), the centroid $d_{k}$ is the average of all the proxy embeddings of $k$th audio, while when calculating positive similarity ($k = n$), we eliminate $e_{mk}$ to compute the centroid.

During the training, the proxy embeddings of each audio should be similar to its centroid, but far from the centroid of other audio in the proxy space. As shown in the similarity matrix in Fig.~\ref{fig:csembedding}, the colored areas should have large values, whereas grey areas should have small values. So for each proxy embedding, the loss function is designed as\vspace{-3pt}
\begin{equation}
L\left({e}_{m n}\right)=-{S}_{m n, n}+\log\left(\sum_{k=1, k\neq n}^{N} \exp \left({S}_{m n, k}\right)\right).\vspace{-3pt}
\end{equation}

After the training of the first stage is completed, we extract the centroid $d_{k}$ of each audio to represent its caption proxy embedding.
\subsubsection{Training details}
In this stage, we use a single-layer LSTM network followed by a linear layer, the dimensions of the LSTM and linear layer are 1024 and 512, respectively. When training the  model, each batch contains $N$ = 64 audio clips and $M$ = 3 captions per audio. The Adam optimizer is used to train the network, the learning rate is 0.01 and the number of training epochs is 500. The scaling factors ($a$,$b$) are initialized as ($10$,$-5$).
\vspace{-8pt}
\subsection{The Second Stage: Regularize The Training of Caption Generation \label{stage2}}
As shown in \textbf{Block B} of Fig.~\ref{fig:acmethod}, the caption proxy embedding $d_k$ is used in the constraint module to regularize the training of the caption generation.

\subsubsection{Training method}We obtain the embedding $\hat{e}_{k}$ of the predicted caption by averaging pooling and max pooling operations on the decoder output representation $H$ along the time axis, as shown in 
\begin{equation}
    \hat{e}_{k} = \mathrm{Mean}(H_k) + \mathrm{Max}(H_k),
\vspace{-5pt}
\end{equation}
where $\hat{e}_{k}$ is the embedding of the $k$th audio predicted caption.

In addition to the cross-entropy loss, we proposed the proxy constraint loss to reduce the effect of the caption disparity (see Eq.~\eqref{eq:scloss}), as shown in 
\begin{equation}
    \ell_{\mathrm{PC}}(\theta  ; d_{k}, x_k) = 1 - \mathrm{cosine}(\hat{e}_{k},d_{k}).
    \label{eq:scloss}
\end{equation}

In this way, a small loss means $\hat{e}_{k}$ has a high similarity with the audio caption proxy embedding $d_{k}$ obtained by the first stage. Accordingly, the final training objective function is the weighted sum of cross-entropy loss and proxy constraint loss, as shown in 
\begin{equation}
 \ell(\theta; d_{k}, C_k, x_k) = \ell_{\mathrm{CE}}(\theta  ; C_k, x_k) + \lambda \cdot \ell_{\mathrm{PC}}(\theta  ; d_{k}, x_k),\vspace{-3pt}
\end{equation}
where $\lambda$ is a hyperparameter.

\begin{table*}[htbp]
\renewcommand{\arraystretch}{1.1}%
\setlength\tabcolsep{3pt}
    \centering
\caption{Experimental results  under Clotho-v1 and Clotho-v2 evaluation sets}
    \vspace{-5pt}
    \begin{tabular}{c|c|cc|ccccccc}\toprule 
         Dataset&Model&Encoder&Decoder&BLEU$_1$&BLEU$_4$&ROUGE$_L$&METEOR&CIDEr&SPICE&SPIDEr \\\midrule 
         \multirow{11}{*}{\rotatebox{90}{Clotho-v1}}&\multirow{4}{*}{Baseline}&\multirow{2}{*}{PANN}&GRU&53.6&13.8&35.9&16.3&34.1&11.0&22.6\\
         & & & Transformer&50.2&11.9&34.0&16.0&32.2&10.7&21.4\\\cline{3-4}
         & &\multirow{2}{*}{Pretrain-CNN}&GRU~\cite{xu2021investigating}&54.3&14.5&36.2&\underline{16.9}&36.5&\textbf{11.7}&24.1\\
         && &Transformer&50.1&12.3&34.6&15.9&31.5&10.7&21.1\\\cline{2-11}
         
         & Fine-tuned PreCNN Transformer~\cite{chen2020audio}&CNN&Transformer&53.4&\underline{15.1}&35.6&16.0&34.6&10.8&22.7\\
         & Temporal attention model~\cite{helinautomated}&CNN&LSTM&48.9&10.7&32.5&14.8&25.2&9.1&17.2\\\cline{2-11} 
         & \multirow{4}{*}{Ours}&\multirow{2}{*}{PANN}&GRU&\underline{55.2}&{15.0}&\underline{36.7}&\textbf{17.1}&\underline{36.9}&\underline{11.6}&\underline{24.2}\\
         & & & Transformer&51.3&13.1&34.9&16.5&33.3&11.2&22.3\\  \cline{3-4}
         & &\multirow{2}{*}{Pretrain-CNN}&GRU&\textbf{55.5}&\textbf{15.3}&\textbf{36.8}&\textbf{17.1}&\textbf{37.5}&\textbf{11.7}&\textbf{24.6}\\
         & & & Transformer&52.4&13.3&35.5&16.6&35.1&11.4&23.2\\ \hline \hline
         
        \multirow{11}{*}{\rotatebox{90}{Clotho-v2}}&\multirow{4}{*}{Baseline}&\multirow{2}{*}{PANN}&GRU&54.5&14.9&36.5&17.0&36.9&11.5&24.2\\
        & & & Transformer&52.0&12.9&35.0&16.8&34.0&11.3&22.6\\\cline{3-4}

        & &\multirow{2}{*}{Pretrain-CNN}&GRU~\cite{xu2021sjtu}&55.2&15.3&37.1&17.3&38.7&11.8&25.2\\
        & & &Transformer&51.7&12.5&34.9&16.5&33.6&11.2&22.4\\\cline{2-11}
        & Transformer+RNN-LM~\cite{narisetty2021leveraging}&PANN&Transformer+RNN-LM&53.3&14.6&35.5&15.4&34.1&10.6&22.4\\
        & CL4AC~\cite{liu2021cl4ac}&PANN&Transformer&55.3&14.3&{37.4}&16.8&36.8&11.5&24.2\\\cline{2-11}
        & \multirow{4}{*}{Ours}&\multirow{2}{*}{PANN}&GRU&55.8&\underline{15.9}&37.4&\underline{17.5}&\underline{39.7}&\underline{12.0}&\underline{25.9}\\

        & & & Transformer&\underline{55.9}&\underline{15.9}&37.3&17.1&37.3&11.6&24.4\\ \cline{3-4}
        & &\multirow{2}{*}{Pretrain-CNN}&GRU&\textbf{56.1}&\textbf{16.0}&\underline{37.5}&\textbf{17.6}&\textbf{40.6}&\textbf{12.1}&\textbf{26.3}\\
        & & & Transformer&\underline{55.9}&\underline{15.9}&\textbf{37.6}&17.2&37.5&11.5&24.5\\\bottomrule
    \end{tabular}
    \label{tab:results}
    \vspace{-18pt}
\end{table*}
\vspace{1pt}
\subsubsection{Training details}In this stage, we use two different 10-layer CNN pre-trained encoder models, PANN\footnote{\url{https://zenodo.org/record/3987831}}
and Pretrain-CNN\footnote{\url{https://zenodo.org/record/5090473}}, and the model structures of them can be found in ~\cite{kong2020panns,xu2021investigating}. And the single-layer attention-based  GRU~\cite{bahdanau2014neural} and 2 layers Transformer~\cite{vaswani2017attention} as our backbone decoder for its success in previous audio captioning works.

SpecAugment~\cite{park2019specaugment} and label smoothing~\cite{szegedy2016rethinking} are applied to prevent overfit. We also apply scheduled sampling to gently decrease the training-inference discrepancy caused by teacher forcing training~\cite{bengio2015scheduled}. The output dimensions of the decoder are 512 and the Transformer decoder has 4 heads. As for the GRU decoder, the Adam optimizer is used to train the network, the initial learning rate is $5\times10^{-4}$ and the total number of training epochs is 25. For the Transformer decoder, the initial learning rate is $5\times10^{-3}$ and warm-up is used to increase the learning rate linearly to the initial learning rate in the first five epochs, the total number of training epochs is 30 and the learning rate is reduced to 1/10 of its original value every 10 epochs. And the hyperparameter $\lambda$ is $0.5$.
\vspace{-5pt}
\section{Experiments}
\vspace{-5pt}
\subsection{Metrics}
We use the same metrics used in the DCASE Challenge to evaluate the performance of our proposed method, including machine translation metrics: BLEU$_n$~\cite{papineni2002bleu},  ROUGE$_L$~\cite{lin2004rouge}, METEOR~\cite{banerjee2005meteor} and captioning metrics: CIDEr~\cite{vedantam2015cider}, SPICE~\cite{anderson2016spice}, SPIDEr~\cite{liu2017improved}. And the $n$ in BLEU means $n-$gram.

The machine translation metrics are used to measure the word accuracy and recall of generated text compared to the ground truth. The captioning metrics are customized to the captioning task, taking into consideration the scene graph contained within the generated caption as well as the $n$-gram's frequency-inverse document frequency (TF-IDF). The semantic fidelity and syntactic fluency of the generated captions are ensured by taking into account the scene graph and the TF-IDF of $n$-gram.

\vspace{-10pt}
\subsection{Datasets}

The Clotho dataset~\cite{drossos2020clotho} and the Audiocaps dataset~\cite{kim2019audiocaps} are two datasets that are commonly used for audio captioning tasks. There are two versions of the Clotho dataset, Clotho-v1 and Clotho-v2 respectively. Each audio clip has five captions describing the audio contents, and the annotator uses only the audio signal for annotation. However, the Audiocaps dataset is not generated in an ideal environment~\cite{kim2019audiocaps}. So in this work, we only focus on the Clotho-v1 and Clotho-v2 datasets.
\vspace{-10pt}
\subsection{Experimental results and discussion}
Tab.~\ref{tab:results} illustrates the experimental results of the proposed method compared to the baseline model and other methods on the Clotho-v1 and Clotho-v2 evaluation sets\footnote{In addition to the metrics, to better verify the
statistical significance of model performance margins, the results of  Student’s t-tests between our methods and the other methods are shown in the Appendix.}. The  \textbf{bold}  and \underline{underline} fonts represent the \textbf{first} and \underline{second} place in each metrics, respectively. 

All baseline methods and our proposed methods are repeated three times, their results in the table are the average of the three experimental results. And the baseline model using Pretrain-CNN encoder and attention-based GRU decoder is our implementation of the state-of-the-art AT-CNN method~\cite{xu2021investigating,xu2021sjtu}. The results of other methods shown in the Tab.~\ref{tab:results} are provided in their paper~\cite{chen2020audio,helinautomated,narisetty2021leveraging,liu2021cl4ac}.

\subsubsection{Compared with baseline models}
For the identical combination of encoder and decoder, our proposed method achieves better results in all evaluation metrics compared to the baseline model. And the baseline model has the identical experimental setting as our proposed method, only without the caption feature space regularization module (\textbf{Block B} of Fig.~\ref{fig:acmethod}). The experimental results and Student’s $t$-tests show that the caption proxy feature space regularization can reduce the effect of caption disparity and significantly improve model performance.

\subsubsection{Compared with other methods} Our proposed best model, in which the encoder is Pretrain-CNN and the decoder is GRU, obtains first or second place in all evaluation criteria in both datasets.
In summary, the proposed method shows outstanding performance on these two datasets.

\vspace{-5pt}
\section{Conclusion}
In this paper, we propose a two-stage caption feature space regularization method for the task of audio captioning. In the proposed method, the first stage uses contrastive learning to generate the proxy feature space and extract the proxy embedding of audio clips. Then the second stage uses the extracted proxy embedding to regularize the training of caption generation and mitigate the effect of the caption disparity. We conducted experiments using four different combinations of pre-trained encoders and decoders on two datasets to demonstrate the effectiveness of the proposed method.

\bibliographystyle{IEEEtran}
\bibliography{ref}

\begin{thebibliography}{10}
\providecommand{\url}[1]{#1}
\csname url@samestyle\endcsname
\providecommand{\newblock}{\relax}
\providecommand{\bibinfo}[2]{#2}
\providecommand{\BIBentrySTDinterwordspacing}{\spaceskip=0pt\relax}
\providecommand{\BIBentryALTinterwordstretchfactor}{4}
\providecommand{\BIBentryALTinterwordspacing}{\spaceskip=\fontdimen2\font plus
\BIBentryALTinterwordstretchfactor\fontdimen3\font minus
  \fontdimen4\font\relax}
\providecommand{\BIBforeignlanguage}[2]{{%
\expandafter\ifx\csname l@#1\endcsname\relax
\typeout{** WARNING: IEEEtran.bst: No hyphenation pattern has been}%
\typeout{** loaded for the language `#1'. Using the pattern for}%
\typeout{** the default language instead.}%
\else
\language=\csname l@#1\endcsname
\fi
#2}}
\providecommand{\BIBdecl}{\relax}
\BIBdecl

\bibitem{drossos2017automated}
K.~Drossos, S.~Adavanne, and T.~Virtanen, ``Automated audio captioning with
  recurrent neural networks,'' in \emph{2017 IEEE Workshop on Applications of
  Signal Processing to Audio and Acoustics (WASPAA)}, 2017, pp. 374--378.

\bibitem{drossos2020clotho}
K.~Drossos, S.~Lipping, and T.~Virtanen, ``Clotho: An audio captioning
  dataset,'' in \emph{ICASSP 2020-2020 IEEE International Conference on
  Acoustics, Speech and Signal Processing (ICASSP)}, 2020, pp. 736--740.

\bibitem{koizumi2020transformer}
Y.~Koizumi, R.~Masumura, K.~Nishida, M.~Yasuda, and S.~Saito, ``A
  transformer-based audio captioning model with keyword estimation,''
  \emph{arXiv preprint arXiv:2007.00222}, 2020.

\bibitem{xu2021audio}
X.~Xu, H.~Dinkel, M.~Wu, and K.~Yu, ``Audio caption in a car setting with a
  sentence-level loss,'' in \emph{2021 12th International Symposium on Chinese
  Spoken Language Processing (ISCSLP)}, 2021, pp. 1--5.

\bibitem{wu2019audio}
M.~Wu, H.~Dinkel, and K.~Yu, ``Audio caption: Listen and tell,'' in
  \emph{ICASSP 2019-2019 IEEE International Conference on Acoustics, Speech and
  Signal Processing (ICASSP)}, 2019, pp. 830--834.

\bibitem{lipping2019crowdsourcing}
S.~Lipping, K.~Drossos, and T.~Virtanen, ``Crowdsourcing a dataset of audio
  captions,'' \emph{arXiv preprint arXiv:1907.09238}, 2019.

\bibitem{rana2016gated}
R.~Rana, ``Gated recurrent unit (gru) for emotion classification from noisy
  speech,'' \emph{arXiv preprint arXiv:1612.07778}, 2016.

\bibitem{Drossos_2019_dcase}
\BIBentryALTinterwordspacing
S.~Lipping, K.~Drossos, and T.~Virtanen, ``Crowdsourcing a dataset of audio
  captions,'' in \emph{Proceedings of the Detection and Classification of
  Acoustic Scenes and Events Workshop ({DCASE})}, Nov. 2019. [Online].
  Available: \url{https://arxiv.org/abs/1907.09238}
\BIBentrySTDinterwordspacing

\bibitem{helinautomated}
B.~Y. Helin~Wang, Y.~Zou, and D.~Chong, ``Automated audio captioning with
  temporal attention.''

\bibitem{chen2020audio}
K.~Chen, Y.~Wu, Z.~Wang, X.~Zhang, F.~Nian, S.~Li, and X.~Shao, ``Audio
  captioning based on transformer and pretrained cnn,'' in \emph{Proceedings of
  the Detection and Classification of Acoustic Scenes and Events Workshop},
  2020, pp. 21--25.

\bibitem{xu2021investigating}
X.~Xu, H.~Dinkel, M.~Wu, Z.~Xie, and K.~Yu, ``Investigating local and global
  information for automated audio captioning with transfer learning,'' in
  \emph{ICASSP 2021-2021 IEEE International Conference on Acoustics, Speech and
  Signal Processing (ICASSP)}, 2021, pp. 905--909.

\bibitem{kim2019audiocaps}
C.~D. Kim, B.~Kim, H.~Lee, and G.~Kim, ``Audiocaps: Generating captions for
  audios in the wild,'' in \emph{Proceedings of the 2019 Conference of the
  North American Chapter of the Association for Computational Linguistics:
  Human Language Technologies, Volume 1 (Long and Short Papers)}, 2019, pp.
  119--132.

\bibitem{eren2020audio}
A.~{\"O}. Eren and M.~Sert, ``Audio captioning based on combined audio and
  semantic embeddings,'' in \emph{2020 IEEE International Symposium on
  Multimedia (ISM)}, 2020, pp. 41--48.

\bibitem{devlin2018bert}
J.~Devlin, M.-W. Chang, K.~Lee, and K.~Toutanova, ``Bert: Pre-training of deep
  bidirectional transformers for language understanding,'' \emph{arXiv preprint
  arXiv:1810.04805}, 2018.

\bibitem{xu2021sjtu}
X.~Xu, Z.~Xie, M.~Wu, and K.~Yu, ``The sjtu system for dcase2021 challenge task
  6: audio captioning based on encoder pre-training and reinforcement
  learning,'' DCASE2021 Challenge, Tech. Rep, Tech. Rep., 2021.

\bibitem{narisetty2021leveraging}
C.~Narisetty, T.~Hayashi, R.~Ishizaki, S.~Watanabe, and K.~Takeda, ``Leveraging
  state-of-the-art asr techniques to audio captioning,'' DCASE2021 Challenge,
  Tech. Rep, Tech. Rep., 2021.

\bibitem{liu2021cl4ac}
X.~Liu, Q.~Huang, X.~Mei, T.~Ko, H.~L. Tang, M.~D. Plumbley, and W.~Wang,
  ``Cl4ac: A contrastive loss for audio captioning,'' \emph{arXiv preprint
  arXiv:2107.09990}, 2021.

\bibitem{kong2020panns}
Q.~Kong, Y.~Cao, T.~Iqbal, Y.~Wang, W.~Wang, and M.~D. Plumbley, ``Panns:
  Large-scale pretrained audio neural networks for audio pattern recognition,''
  \emph{IEEE/ACM Transactions on Audio, Speech, and Language Processing},
  vol.~28, pp. 2880--2894, 2020.

\bibitem{bahdanau2014neural}
D.~Bahdanau, K.~Cho, and Y.~Bengio, ``Neural machine translation by jointly
  learning to align and translate,'' \emph{arXiv preprint arXiv:1409.0473},
  2014.

\bibitem{vaswani2017attention}
A.~Vaswani, N.~Shazeer, N.~Parmar, J.~Uszkoreit, L.~Jones, A.~N. Gomez,
  {\L}.~Kaiser, and I.~Polosukhin, ``Attention is all you need,''
  \emph{Advances in neural information processing systems}, vol.~30, 2017.

\bibitem{park2019specaugment}
D.~S. Park, W.~Chan, Y.~Zhang, C.-C. Chiu, B.~Zoph, E.~D. Cubuk, and Q.~V. Le,
  ``Specaugment: A simple data augmentation method for automatic speech
  recognition,'' \emph{arXiv preprint arXiv:1904.08779}, 2019.

\bibitem{szegedy2016rethinking}
C.~Szegedy, V.~Vanhoucke, S.~Ioffe, J.~Shlens, and Z.~Wojna, ``Rethinking the
  inception architecture for computer vision,'' in \emph{Proceedings of the
  IEEE conference on computer vision and pattern recognition}, 2016, pp.
  2818--2826.

\bibitem{bengio2015scheduled}
S.~Bengio, O.~Vinyals, N.~Jaitly, and N.~Shazeer, ``Scheduled sampling for
  sequence prediction with recurrent neural networks,'' \emph{Advances in
  neural information processing systems}, vol.~28, 2015.

\bibitem{papineni2002bleu}
K.~Papineni, S.~Roukos, T.~Ward, and W.-J. Zhu, ``Bleu: a method for automatic
  evaluation of machine translation,'' in \emph{Proceedings of the 40th annual
  meeting of the Association for Computational Linguistics}, 2002, pp.
  311--318.

\bibitem{lin2004rouge}
C.-Y. Lin, ``Rouge: A package for automatic evaluation of summaries,'' in
  \emph{Text summarization branches out}, 2004, pp. 74--81.

\bibitem{banerjee2005meteor}
S.~Banerjee and A.~Lavie, ``Meteor: An automatic metric for mt evaluation with
  improved correlation with human judgments,'' in \emph{Proceedings of the acl
  workshop on intrinsic and extrinsic evaluation measures for machine
  translation and/or summarization}, 2005, pp. 65--72.

\bibitem{vedantam2015cider}
R.~Vedantam, C.~Lawrence~Zitnick, and D.~Parikh, ``Cider: Consensus-based image
  description evaluation,'' in \emph{Proceedings of the IEEE conference on
  computer vision and pattern recognition}, 2015, pp. 4566--4575.

\bibitem{anderson2016spice}
P.~Anderson, B.~Fernando, M.~Johnson, and S.~Gould, ``Spice: Semantic
  propositional image caption evaluation,'' in \emph{European conference on
  computer vision}.\hskip 1em plus 0.5em minus 0.4em\relax Springer, 2016, pp.
  382--398.

\bibitem{liu2017improved}
S.~Liu, Z.~Zhu, N.~Ye, S.~Guadarrama, and K.~Murphy, ``Improved image
  captioning via policy gradient optimization of spider,'' in \emph{Proceedings
  of the IEEE international conference on computer vision}, 2017, pp. 873--881.

\end{thebibliography}
\vspace{-0.5cm}

\end{document}


\appendices
\section{Results of Student's $t$-tests}
To better verify the statistical significance of model performance margins, here list the $p$-values of Student’s \emph{t}-tests between our methods and state-of-the-art methods as supplements. Specifically, we conduct two-sample Student’s \emph{t}-tests with the null hypothesis that the means of two populations are equal. With the significance level set as $0.05$, the performance margin is statistically significant when its corresponding $p$-value is smaller than $0.05$. According to the results of Student’s \emph{t}-tests shown in Table~\ref{t-test_baseline} and Table~\ref{t-test_other}, respectively, we can demonstrate that (i) our proposed methods can obtain significant improvement compared to the baseline methods with the same combination of encoder and decoder(\emph{i.e.}, outperform baseline methods with $p$-value smaller than $0.05$), and (ii) our proposed best model, which uses Pretrain-CNN as the encoder and GRU as the decoder, also can achieve statistically significant differences in most of the metrics with other methods in both datasets.

\begin{table}[htbp]
\renewcommand{\arraystretch}{1.1}%
\caption{Student’s \emph{t}-tests between our proposed method and the same configuration of the baseline method on both datasets. The significance level was set as $0.05$. "$\surd$" indicates the null hypothesis is rejected, and "$\times$" means the null hypothesis is accepted.}
\label{t-test_baseline}
\centering
    \begin{tabular}{c|cc|ccccccc}\toprule 
     Dataset&Encoder &Decoder&BLEU$_1$&BLEU$_4$&ROUGE$_L$&METEOR&CIDEr&SPICE&SPIDEr \\\midrule 
     \multirow{4}{*}{{Clotho-v1}}&PANN &GRU&$\surd$&$\surd$&$\surd$&$\surd$&$\surd$&$\surd$&$\surd$\\
     &PANN & Transformer&$\times$&$\surd$&$\surd$&$\surd$&$\surd$&$\surd$&$\surd$\\
     &Pretrain-CNN &GRU&$\surd$&$\surd$&$\surd$&$\surd$&$\surd$&$\times$&$\surd$\\
     &Pretrain-CNN &Transformer&$\surd$&$\surd$&$\surd$&$\times$&$\surd$&$\surd$&$\surd$\\\hline\hline
    \multirow{4}{*}{{Clotho-v2}}&PANN & GRU&$\surd$&$\surd$&$\surd$&$\surd$&$\surd$&$\surd$&$\surd$\\
     &PANN & Transformer&$\surd$&$\surd$&$\surd$&$\times$&$\surd$&$\times$&$\surd$\\
     &Pretrain-CNN & GRU&$\times$&$\surd$&$\surd$&$\surd$&$\surd$&$\surd$&$\surd$\\
     &Pretrain-CNN & Transformer&$\surd$&$\surd$&$\surd$&$\surd$&$\surd$&$\times$&$\surd$\\\bottomrule
    \end{tabular}
\end{table}

\begin{table}[htbp]
\renewcommand{\arraystretch}{1.1}%
\caption{Student’s \emph{t}-tests between our proposed best method and other methods. The significance level was set as $0.05$. "$\surd$" indicates the null hypothesis is rejected, and "$\times$" means the null hypothesis is accepted.}
\label{t-test_other}
\centering
    \begin{tabular}{c|c|ccccccc}\toprule 
     Dataset&Method&BLEU$_1$&BLEU$_4$&ROUGE$_L$&METEOR&CIDEr&SPICE&SPIDEr \\\midrule 
     \multirow{2}{*}{{Clotho-v1}}&Fine-tuned PreCNN Transformer&$\surd$&$\times$&$\surd$&$\surd$&$\surd$&$\surd$&$\surd$\\
     &Temporal attention model&$\surd$&$\surd$&$\surd$&$\surd$&$\surd$&$\surd$&$\surd$\\\hline\hline
    \multirow{2}{*}{{Clotho-v2}}&Transformer+RNN-LM&$\surd$&$\surd$&$\surd$&$\surd$&$\surd$&$\surd$&$\surd$\\
     &CL4AC&$\times$&$\surd$&$\times$&$\surd$&$\surd$&$\surd$&$\surd$\\\bottomrule
    \end{tabular}
\end{table}